\documentclass[preprintnumbers,showpacs,floats,twocolumn,prd,aps]{revtex4}
\usepackage{amssymb,amsmath}
\usepackage{epsfig,hyperref}
\usepackage[svgnames]{xcolor}
\usepackage{pgf,tikz}
\usepackage{epstopdf}

\def\be{\begin{equation}}
\def\ee{\end{equation}}
\def\beq{\begin{eqnarray}}
\def\eeq{\end{eqnarray}}

\makeatletter
\renewcommand{\vec}[1]{\mbox{\boldmath$#1$}}
\newcommand{\arXiv}[2][]{\href{http://arxiv.org/abs/#2}{\texttt{arXiv:#2\@ifempty{#1}{}{ [#1]}}}}
\makeatother

\begin{document}


\title{Construction of $\boldsymbol{f(R)}$ Gravity Models}

\author{Jun-Qi Guo}%
\email{jga35@sfu.ca}
\affiliation{%
 Department of Physics, Simon Fraser University\\
8888 University Drive, Burnaby, BC Canada V5A 1S6
}%
\date{July 11, 2013}

\begin{abstract}
In this paper, we study how to construct $f(R)$ gravity models.
Cosmological observations and local gravity tests imply that a viable $f(R)$ model should be very close to the $\Lambda$CDM model. We create procedures to construct viable $\Lambda$CDM-like $f(R)$ models, and present multiple models of three types. The connections between some of these models are discussed.
We also study the cosmological evolution of the $\Lambda$CDM-like $f(R)$ gravity. Exact numerical integration can generate accurate cosmological evolution, but the numerical simulation in the early-universe stage is slow due to the oscillations of the field $\phi(\equiv f')$ near the minimum of the effective potential $V_{\text{eff}}(\phi)$. To avoid this problem, we take the minimum of the effective potential $V_{\text{eff}}(\phi)$ as an approximate solution for $\phi$, and obtain the cosmological evolution. This approximate method describes the cosmological evolution well except in the late universe. Therefore, we use the approximate method in the early-universe evolution, and use the exact method in the late-universe one.

\end{abstract}

\preprint{SCG-2013-06}
\maketitle

\section{Introduction\label{sec:introduction}}
The causes of cosmological acceleration remain unknown \cite{Riess:1998cb, Perlmutter:1998np, Riess:2004nr, Komatsu:2010fb, Planck:2013}. Among various approaches to explain this cosmic speed-up, $f(R)$ gravity is a straightforward option. In the Jordan frame, one may replace the Ricci curvature scalar in the Einstein-Hilbert action with a function of the scalar
\be S_{\text{JF}}=\frac{1}{16\pi G}\int d^{4}x \sqrt{-g}f(R)+S_M\left[g_{\mu\nu},\psi_{m}\right], \label{f_R_action} \ee
where $G$ is the Newtonian gravitational constant, and $\psi_{m}$ is the matter field. See Refs.~\cite{Sotiriou,Tsujikawa1} for reviews of $f(R)$ theory.

The $f(R)$ gravity needs to confront stability, cosmological viability, and local gravity tests. In fact, the features of an $f(R)$ model are largely determined by a potential $V(\phi)$ as defined in the equations of motion for $f(R)$ theory. In this paper, we explore how to construct viable $f(R)$ models by connecting the viability conditions with the geometry of the potential. Cosmological observations and local gravity tests imply that a viable $f(R)$ model should be very close to the $\Lambda$CDM model. Here we make instructions to build viable $\Lambda$CDM-like $f(R)$ models, and present multiple models (three types). Some of them have been proposed before, but others have not. We also point out the connections between some of these models.

Cosmological evolution and solar system tests are two extreme cases of the dynamics of the field $\phi\equiv f'$. In the cosmological evolution, the universe is assumed to be homogeneous. Therefore, the cosmic dynamics only depend on the temporal variable. In the solar system tests, the spherically symmetric spacetime is static, then the field $\phi$ only varies with respect to the spacial variable. In both cases, the field $\phi$ interacts with the matter density, and the field $\phi$ can be coupled to the matter density when the matter density is much greater than the cosmological constant. The solar system tests of $f(R)$ gravity are explored by considering an effective potential constructed by the matter density and the
potential $V(\phi)$~\cite{Justin1,Justin2,Navarro,Faulkner,Hu_Sawicki,Gu,Tsujikawa2,Tsujikawa3,Guo1}. In this paper, we employ the same approach to study the cosmic dynamics of $f(R)$ gravity.

For a feasible $f(R)$ model, in the early universe, the field $\phi(\equiv f')$ has a slow roll evolution so that a matter domination phase exists. In the late universe, the field $\phi$ will be released from the coupling between the matter density and the potential $V(\phi)$, thus generating the cosmic speed-up. In this study, the cosmic dynamics of $f(R)$ gravity are explored from the early universe to the late one. Due to the oscillations of $\phi$ in the effective potential $V_{\text{eff}}(\phi)$, the numerical simulation of the cosmological evolution can be very slow. In order to integrate the evolution more effectively, the evolution of the minimum point of $V_{\text{eff}}(\phi)$ is taken as an approximate solution for $\phi$. This approach describes the cosmological evolution well until in the late universe. To supplement, the exact method is used in the late universe. Therefore, a combination of exact and approximate methods provides a complete picture of the cosmological evolution of $f(R)$ gravity.

The paper is organized as follows.
In Sec.~\ref{sec:framework}, we introduce the framework.
In Sec.~\ref{sec:class_I_models}, a type of non-$\Lambda$CDM-like $f(R)$ model will be explored  .
Sec.~\ref{sec:LCDM_like_models} discusses how to construct viable $\Lambda$CDM-like $f(R)$ models.
In Sec.~\ref{sec:cosmological_evolution}, the cosmological evolution of one example of these $\Lambda$CDM-like models is analysed.
Lastly, Sec.~\ref{sec:conclusions} summarizes our results.
\section{Framework\label{sec:framework}}
%
A variation on the action for $f(R)$ gravity with respect to the metric yields gravitational equations of motion
\be f'R_{\mu\nu}-\frac{1}{2}f g_{\mu\nu} -[\nabla_{\mu}\nabla_{\nu}-g_{\mu\nu} \Box] f'
= 8\pi GT_{\mu\nu},\label{gravi_eq_fR} \ee
where $f'$ denotes the derivative of the function $f$ with respect to its argument $R$, and $\Box$ is the usual
notation for the covariant D'Alembert operator $\Box\equiv\nabla_{\alpha}\nabla^{\alpha}$. Compared to general relativity, $f(R)$ gravity has one extra scalar degree of freedom, $f'$. The dynamics of this degree of freedom are determined by the trace of Eq. (\ref{gravi_eq_fR}),
\be \Box f'=\frac{1}{3}(2f-f'R) + \frac{8\pi G}{3}T, \label{trace_eq1}\ee
where $T$ is the trace of the stress-energy tensor $T_{\mu\nu}$. Identifying $f'$ by
\be \phi\equiv\frac{df}{dR}, \label{f_prime}\ee
and defining a potential $V(\phi)$ by
\be V'(\phi)\equiv\frac{dV}{d\phi}=\frac{1}{3}\left(2f-\phi R\right), \label{v_prime} \ee
one can rewrite Eq.~(\ref{trace_eq1}) as
\be \Box \phi=V'(\phi)+\frac{8\pi G}{3}T.\label{trace_eq2}\ee
In order to explore how $f(R)$ gravity causes cosmic speed-up, it is convenient to cast the formulation of $f(R)$
gravity in a format similar to that of general relativity. We can rewrite Eq.~(\ref{gravi_eq_fR}) as
\be G_{\mu \nu}=8\pi G \left[ T_{\mu \nu} + T_{\mu \nu} ^{(\text{eff})} \right],
\label{field_eq4} \ee
where
\be 8\pi GT_{\mu \nu} ^{(\text{eff})}=\frac{f-f'R}{2}g_{\mu\nu}+\nabla_{\mu}\nabla_{\nu}f'-g_{\mu\nu}\Box f' + (1-f')G_{\mu\nu}. \label{tilde_T4}  \ee
$T^{\mu\nu}_{\text{(eff)}}$ is the energy-momentum tensor of the effective dark energy, and it is guaranteed to be conserved, $\tilde{T}^{\mu\nu}_{\text{(eff)};\nu} = 0$. Equation~(\ref{tilde_T4}) gives the definition of the equation of state for the effective dark energy as
\be w_{\text{eff}}\equiv\frac{p_{\text{eff}}}{\rho_{\text{eff}}}, \label{w_eff}\ee
where
\be
\begin{split}
8\pi G \rho_{\text{eff}}& =3H^{2}-8\pi G(\rho_m+\rho_r)\\
& =\frac{f'R-f}{2}-3H\dot{f'} +3H^{2}(1-f'),
\end{split}
\label{rho_eff}
\ee
\be
\begin{split}
8\pi G p_{\text{eff}}& =H^{2}-R/3-8\pi G p_r\\
& =\ddot{f'}+2H\dot{f'}+\frac{f-f'R}{2}+(H^{2}-R/3)(1-f').
\end{split}
\label{p_eff}
\ee
In order for an $f(R)$ model to account for the cosmic speed-up, $w_{\text{eff}}$ should be less than $-1/3$.
\section{Non-$\boldsymbol{\Lambda}$CDM-like $\boldsymbol{f(R)}$ models\label{sec:class_I_models}}

\subsection{Viability conditions on $\boldsymbol{f(R)}$ gravity\label{sec:viability_conditions}}
A viable $f(R)$ model should be stable, mimic a cosmological evolution consistent with observations, and satisfy local gravity tests. This places some viability conditions on $f(R)$ gravity as follows.
\begin{enumerate}
\item We require $f'$ to be positive to avoid anti-gravity.
\item The function $f(R)$ should be very close to the curvature scalar $R$ at high curvature so that a matter domination epoch can exist in the early universe.
\item The $f''$ should be positive when the curvature scalar $R$ is greater than the cosmological constant $\Lambda$, so that the Dolgov-Kawasaki instability can be avoided and the scalaron $f'$ is non-tachyonic~\cite{Dolgov}. Moreover, the potential $V(\phi)$ should have a minimum such that a dark-energy domination stage and a consequent cosmic acceleration can be generated in the late universe.
\item The Big Bang nucleosynthesis, observations of the Cosmic Microwave Background, and local gravity tests imply that general relativity should be recovered as $R\gg\Lambda$: $f(R)\rightarrow R$ and $f'\rightarrow 1$. This, together with the requirement of $f''>0$, implies that $f'$ should be less than 1~\cite{Pogosian}.
\end{enumerate}
\subsection{Non-$\boldsymbol{\Lambda}$CDM-like $\boldsymbol{f(R)}$ models}
\begin{figure}[!htbp]
\includegraphics[width=7.5cm,height=5.8cm]{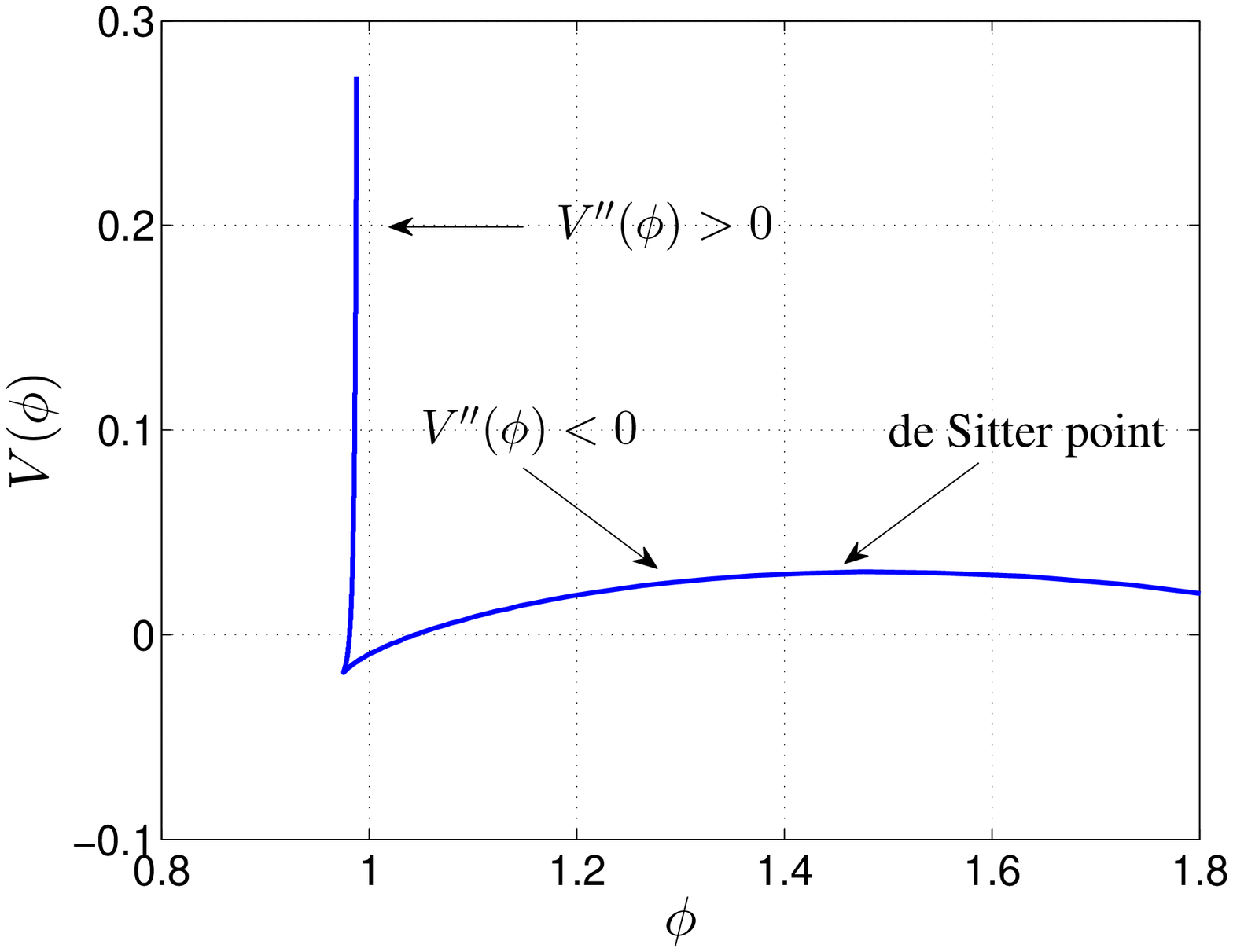}
\caption{The potential for the modified logarithmic model of Eq.~(\ref{modified_log}) with $b=0.1$ and $R_0=1$. The potential $V(\phi)$ is folded before $\phi$ reaches the de Sitter point. At the folding point, $V''(\phi)$ switches signs. The model is unstable at places where $V''(\phi)<0$.} \label{fig:potential_small_b}
%
\includegraphics[width=7.5cm,height=5.8cm]{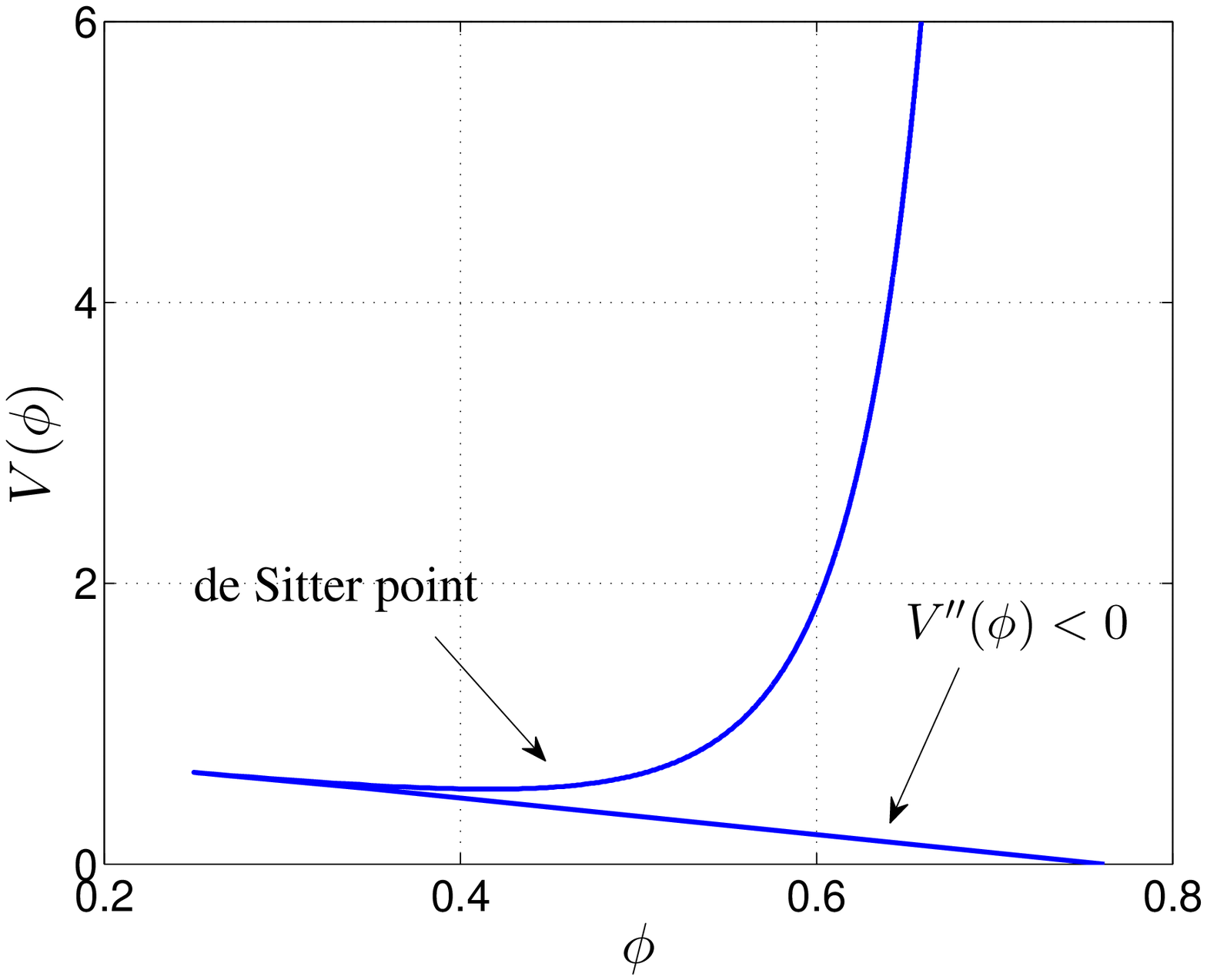}
\caption{The potential for the modified logarithmic model of Eq.~(\ref{modified_log}) with $b=3$ and $R_0=1$. The folding point of $V(\phi)$ is shifted to the left side of the de Sitter point in this configuration.} \label{fig:potential_large_b}
\end{figure}

In our previous work~\cite{Frolov,Guo1,Guo2}, we explored an $R\ln R$ model, $f(R)=R\left[1+\alpha\ln (R/R_0)\right]$. In this model, the modification term causes significant deviation from general relativity at high curvature. As a result, this model has difficulties when it comes to developing a matter domination stage and passing solar system tests. This problem is alleviated in the modified logarithmic model
\be f(R)=R\frac{a+\log (R/R_0)}{1+\log (R/R_0)}=R\left[1-\frac{b}{1+\log (R/R_0)}\right], \label{modified_log}\ee
where $b=1-a$. In this model,
\be f'=1-\frac{b}{1+\log (R/R_0)}+\frac{b}{[1+\log (R/R_0)]^{2}}.\label{f_prime_modified_log}\ee
Equation~(\ref{f_prime_modified_log}) implies that $f''$ and $V''(\phi)(=(f'-f''R)/3f'')$ will change signs at some point, and the potential $V(\phi)$ is folded at that point. When $b$ is small, the potential $V(\phi)$ can be folded before $\phi$ reaches the de Sitter point, as shown in Fig.~\ref{fig:potential_small_b}. The model is unstable at places where $f''<0$ and $V''(\phi)<0$. When $b$ is greater, the folding point of $V(\phi)$ can be shifted to the left side of the de Sitter point, as shown in Fig.~\ref{fig:potential_large_b}. Because of the \lq\lq soft\rq\rq~logarithmic dependence in the function $f(R)$, this model does not have a fully matter-dominated epoch in the early universe.
\begin{figure}
\includegraphics[width=7.5cm,height=5.8cm]{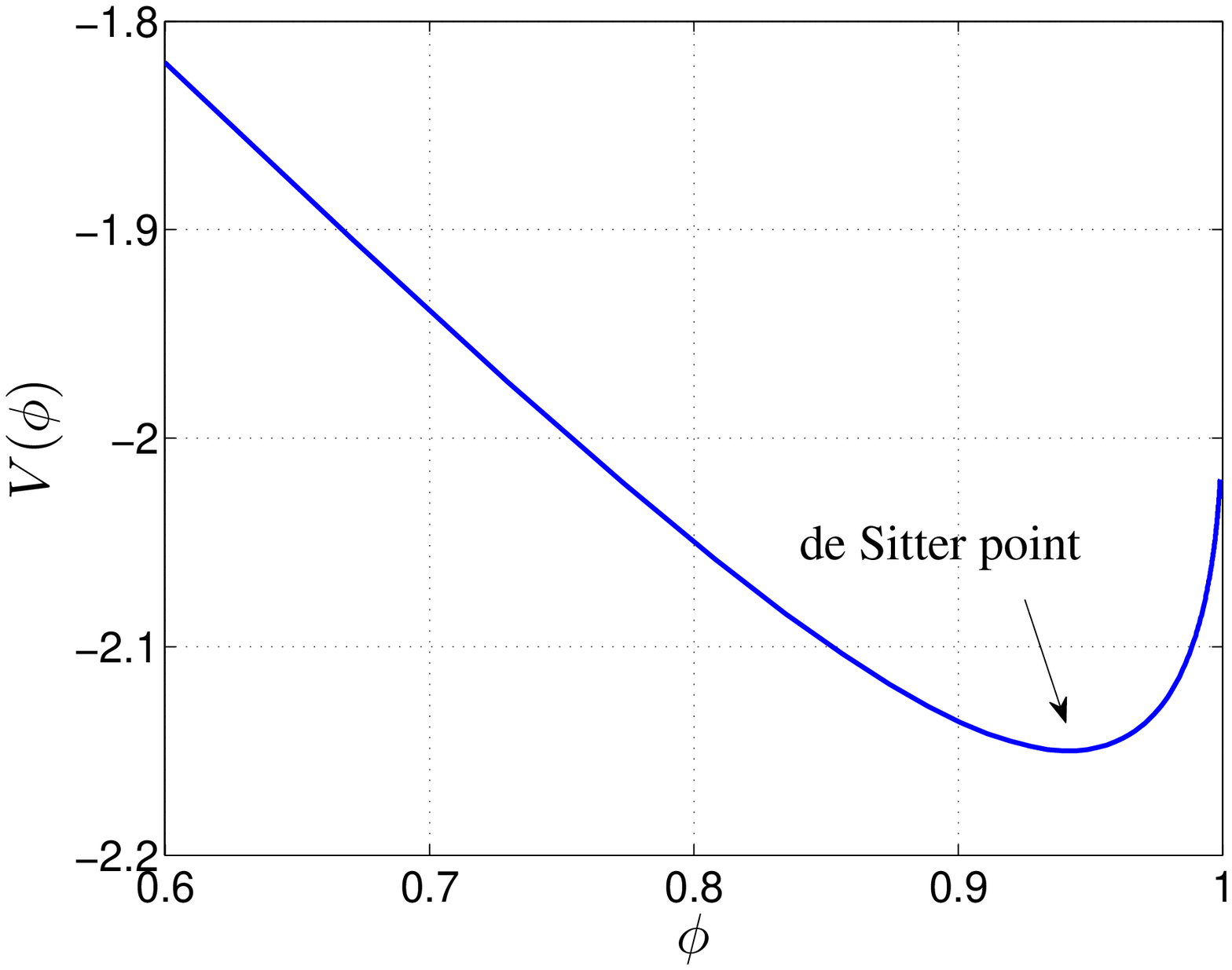}
\caption{The potential for the polynomial model of Eq.~(\ref{class_I_polynomial}) with $n=R_0=1$ and $b=5$. In the case of $n=1$, this model happens to be the simplest format of the Hu-Sawicki model of Eq.~(\ref{Hu_Sawicki_model}), $V(\phi)$ does not have a folding point and $V''(\phi)>0$ from the high ($R\gg \Lambda$) to the low ($R\sim \Lambda$) curvature regimes.}
\label{fig:potential_polynomial_n_1}
%
\includegraphics[width=7.5cm,height=5.8cm]{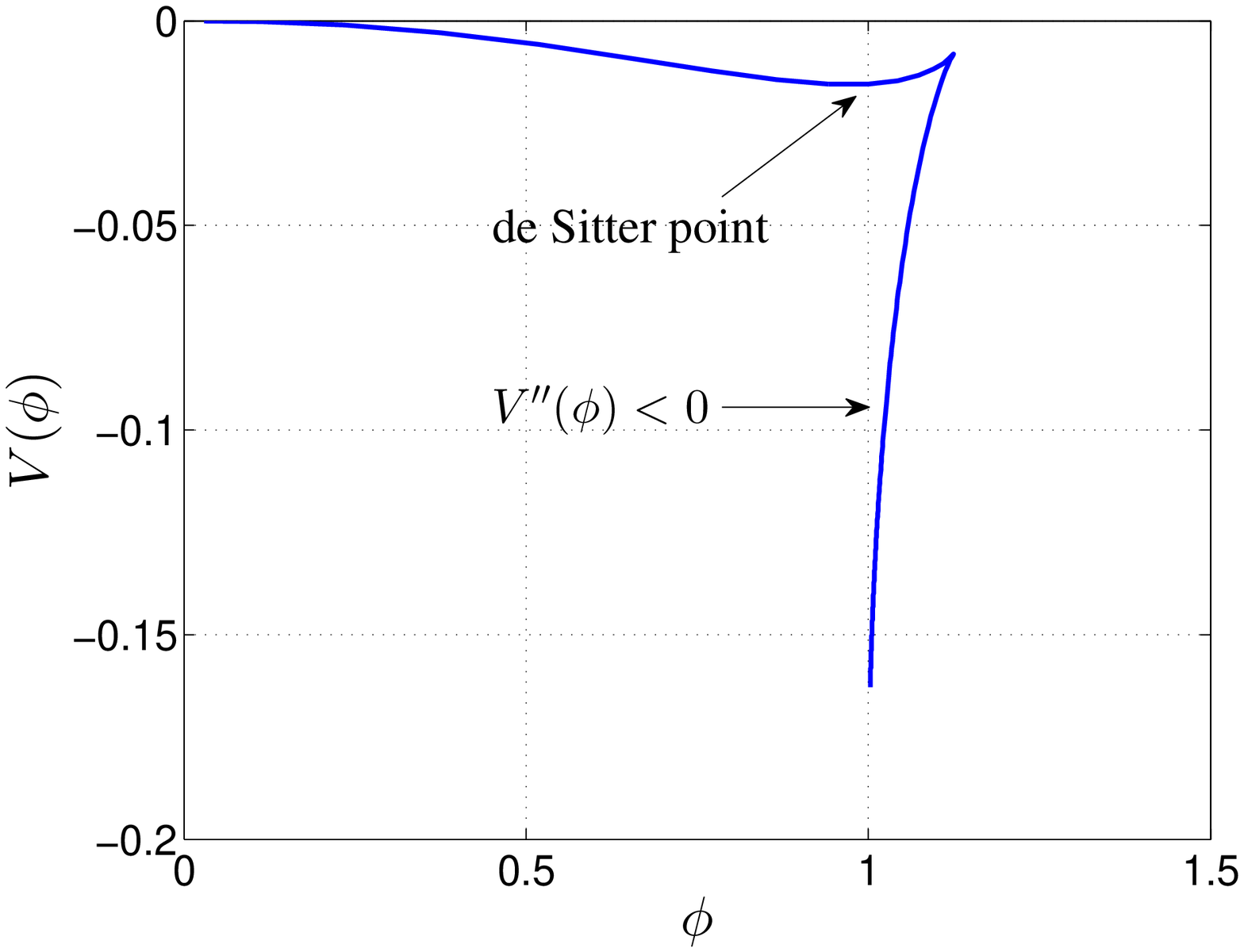}
\caption{The potential for the polynomial model of Eq.~(\ref{class_I_polynomial}) with $n=2$ and $b=R_0=1$. $V(\phi)$ has a folding point. For some period of $\phi$ from the high ($\phi\rightarrow 1$) to the low curvature regimes, $V''(\phi)$ is negative and, consequently, the scalaron $f'$ is tachyonic.}
\label{fig:potential_polynomial_n_2}
\end{figure}
\begin{figure}
\includegraphics[width=7.5cm,height=5.8cm]{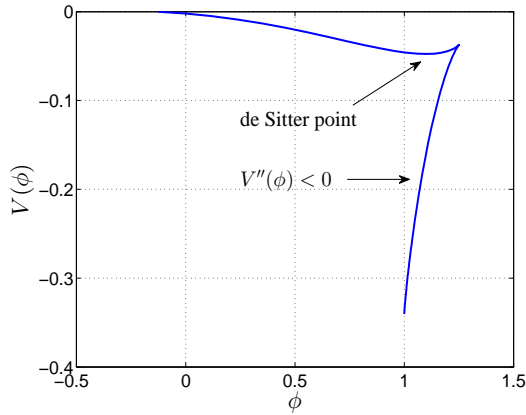}
\caption{The potential for the exponential model of Eq.~(\ref{class_I_exponential}) with $b=2.5$ and $R_0=1$. $V(\phi)$ has a folding point. For some period of $\phi$ from the high ($\phi\rightarrow 1$) to the low curvature regimes, $V''(\phi)$ is negative and, consequently, the scalaron $f'$ is tachyonic.}
\label{fig:potential_exp_model}
\end{figure}
In order to explore other possibilities, one can generalize the function of Eq.~(\ref{modified_log}) as follows:
\be f(R)=R\left[1-\frac{b}{1+A(R/R_0)}\right], \label{class_I_models}\ee
where $A(R/R_0)$ is a function of $R/R_0$. One may consider a polynomial case as expressed by
\be f(R)=R\left[1-\frac{b}{1+(R/R_0)^{n}}\right]. \label{class_I_polynomial}\ee
When $n=1$, this model happens to be the simplest format of the Hu-Sawicki model~\cite{Hu_Sawicki}. The complete format of the Hu-Sawicki model is described by Eq.~(\ref{Hu_Sawicki_model}). In the case of $n=1$,
\be f'=1-\frac{b}{(1+R/R_0)^2}.\ee
Therefore, $f''$ will not be zero, and $V(\phi)$ does not have a folding point from the high curvature regime ($R\gg \Lambda$) to the low one ($R\sim \Lambda$), as is shown in Fig.~\ref{fig:potential_polynomial_n_1}.
However, as long as $n>1$, $f''$ will include more than one term of $R$ with different signs, $V(\phi)$ will have a folding point, and $V''(\phi)$ will switch signs at the folding point. As illustrated in Fig.~\ref{fig:potential_polynomial_n_2}, in this case, for some period of $\phi$ from the high to the low curvature regimes, $V''(\phi)$ is negative and, consequently, the scalaron $f'$ is tachyonic.
We also considered the exponential case
\be f(R)=R\left[1-\frac{b}{1+\exp(R/R_0)}\right]. \label{class_I_exponential}\ee
This model has a similar problem as shown in Fig.~\ref{fig:potential_exp_model}.

In this class of $f(R)$ models of Eq.~(\ref{class_I_models}), there are functions of $R$ in both the numerator and the denominator of the modification term, which results in $V(\phi)$ having a folding point. At that folding point, $V''(\phi)$ switches signs, and the scalaron $f'$ is tachyonic at places where $V''(\phi)<0$. To avoid this problem, one may replace the function of $R$ in the numerator with a constant, and obtain $\Lambda$CDM-like models. In these models, the function $f(R)$ is approximately equal to $(R-2\Lambda)$ at high curvature, and the modification term will be the decisive term in the late universe.
\section{Construction of $\boldsymbol{\Lambda}$CDM-like $\boldsymbol{f(R)}$ models\label{sec:LCDM_like_models}}
\subsection{Procedures\label{sec:procedures}}
In this study, we construct three types of $\Lambda$CDM-like $f(R)$ models as expressed by Eqs.~(\ref{Type_I_LCDM_models}), (\ref{Type_II_LCDM_models}), and (\ref{Type_III_LCDM_models}), respectively. For simplicity, we take the Type II $f(R)$ gravity described by Eq.~(\ref{Type_II_LCDM_models}) as an example to explain how to construct viable $\Lambda$CDM-like $f(R)$ models. The function for the Type II $f(R)$ gravity is
\be f(R)=R-b[c-A(R/R_0)]. \ee
The procedures are as follows.
\begin{enumerate}
\item The parameters $b$ and $R_0$ have the same energy scale as the cosmological constant. For simplicity, $R_0$ is set to be equal to 1.
\item Generally, $A(R/R_0)$ goes to zero at high curvature, such that the $f(R)$ model reduces to the $\Lambda$CDM model at high curvature. This will make the $f(R)$ model have a matter domination epoch in the early universe, and avoid the solar system tests as well.
\item At high curvature, $f'$ should be positive to avoid anti-gravity. This can be guaranteed without difficulty, because at high curvature regime the modification term $b[c+A(R/R_0)]$ is much less than the main term $R$ in the function $f(R)$.
\item The $f''$ should be positive when $R>\Lambda$, such that the Dolgov-Kawasaki instability can be avoided ~\cite{Dolgov} and the scalaron $f'$ is non-tachyonic. The potential $V(\phi)$ should have a minimum, such that the model is stable and can mimic the later cosmic acceleration. These can be obtained by the following measures.
    \begin{enumerate}
      \item Note that in $f(R)$ theory
      \be V''(\phi)=\frac{f'-f''R}{3f''}.\ee
      Generally, $f'>f''R$ and $f'\sim 1$ for the $\Lambda$CDM-like models. Therefore, $V''(\phi)$ is mainly determined by $f''$, and $f''$ is determined by the modification term in the function $f(R)$. Thus we make sure that $f''>0$ and $V''>0$ by tuning the sign before $A(R/R_0)$.
      \item Let $V'(\phi)=(2f-f'R)/3>0$ at high curvature. This can be trivially satisfied for a $\Lambda$CDM-like model for which $f(R)\rightarrow R$ and $f'\rightarrow 1$ at high curvature.
      \item Tune the parameters $b$ and $c$ to make sure that $V'(\phi)=(2f-f'R)/3<0$ as $R\sim \Lambda$.
   \end{enumerate}
\item The requirement of $f'<1$ will be consequently satisfied for the $\Lambda$CDM-like models once the requirement of $f''>0$ is met.
\end{enumerate}

\subsection{Type I $\boldsymbol{\Lambda}$CDM-like $\boldsymbol{f(R)}$ models}
One obtains the first type of $\Lambda$CDM-like $f(R)$ model by replacing the term $b\cdot R$ in the numerator of the modification term in Eq.~(\ref{class_I_models}) with the parameter $b$, so that
\be f(R)=R-\frac{b}{c+A(R/R_0)}, \label{Type_I_LCDM_models}\ee
where $c$ is another parameter. One can construct some viable models as follows by letting $A(R/R_0)$ take some elementary functions and implementing the procedures discussed in Sec.~\ref{sec:procedures}.

\begin{figure}
\includegraphics[width=7.5cm,height=5.8cm]{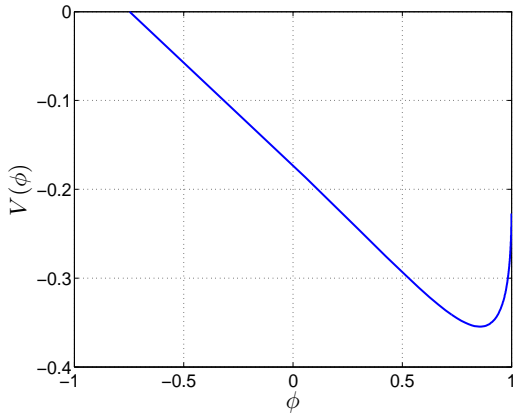}
\caption{The potential for the model of Eq.~(\ref{log_model_class_2_1}) with $b=1.5$ and $c=R_0=1$. The potential has a minimum, and $V''(\phi)$ is positive from the high ($R\gg \Lambda$, $\phi\rightarrow 1$) to the low ($R\sim\Lambda$, $\phi\sim 0.8$) curvature regimes. Therefore, this model should be stable, and should also have a sensible cosmological evolution as verified in Sec.~{\ref{sec:cosmological_evolution}}.}
\label{fig:potential_log_model}
\end{figure}

\vspace{3mm}
\noindent\textbullet{ \emph{Logarithmic format 1:}}
\be f(R)=R-\frac{b}{c+\log(1+R_0/R)}. \label{log_model_class_2_1}\ee
Example parameters for this model are $b=1.5$ and $c=R_0=1$. The potential for this model with these parameters is plotted in Fig.~\ref{fig:potential_log_model}. As expected, the potential has a minimum, and $V''(\phi)$ is positive from the high ($R\gg \Lambda$, $\phi\rightarrow 1$) to the low ($R\sim\Lambda$, $\phi\sim 0.8$) curvature regimes. Therefore, this model should be stable, and should also have a sensible cosmological evolution as verified in Sec.~{\ref{sec:cosmological_evolution}}. The potentials for all the $\Lambda$CDM-like $f(R)$ models presented in this paper have been plotted, with the parameters taking appropriate values. They all look similar to Fig.~\ref{fig:potential_log_model} and therefore are not individually shown here.

\vspace{3mm}
\noindent\textbullet{ \emph{Logarithmic format 2:}}
\be f(R)=R-\frac{b}{c+1/\log(1+R/R_0)}. \ee
For this model, example parameters, which can generate a potential similar to the one plotted in Fig.~\ref{fig:potential_log_model}, are $b=c=5$ and $R_0=1$.

\vspace{3mm}
\noindent\textbullet{ \emph{Polynomial format:}}
\be f(R)=R-\frac{b}{c+(R_0/R)^{n}}=R-\frac{b(R/R_0)^{n}}{c(R/R_0)^{n}+1}, \label{Hu_Sawicki_model}\ee
where $n$ is a positive integer number. Example parameters for this model are $b=2$, $c=R_0=1$, and $n=3$. This is the Hu-Sawicki model~\cite{Hu_Sawicki}.

\vspace{3mm}
\noindent\textbullet{ \emph{Exponential format 1:}}
\be f(R)=R-\frac{b}{c+\exp(-R/R_0)}. \ee
Example parameters for this model are $b=2$ and $c=R_0=1$. This model is almost the same as the one discussed in Ref.~\cite{Oikonomou}:
\be f(R)=R-\frac{C}{A+B\exp(-R/D)}+\frac{C}{A+B},\label{Type_I_exp_Oikonomou}\ee
where $A, B, C,$ and $D$ are parameters.

\vspace{3mm}
\noindent\textbullet{ \emph{Exponential format 2:}}
\be f(R)=R-\frac{b}{c+\exp(R_0/R)}. \ee
Example parameters for this model are $b=5$, and $c=R_0=1$.
\subsection{Type II $\boldsymbol{\Lambda}$CDM-like $\boldsymbol{f(R)}$ models}
In the Type I models described by Eq.~(\ref{Type_I_LCDM_models}), the modification term, $c+A(R/R0)$, is always finite. Therefore, we can move it to the numerator with the sign before $A(R/R0)$ switched, so that
\be f(R)=R-b[c-A(R/R_0)]. \label{Type_II_LCDM_models}\ee
Some models of this type can be constructed as follows.

\vspace{3mm}
\noindent\textbullet{ \emph{Logarithmic format 1:}}
\be f(R)=R-b\left[c-\log\left(1+\frac{R_0}{R}\right)\right].\label{log_model_class_3_1}\ee
Example parameters for this model are $b=c=R_0=1$. When $c$ is equal to zero, one obtains
\be f(R)=R-\alpha R_0\log\left(1+\frac{R}{R_0}\right),\label{log_model_class_3_2}\ee
where $\alpha$ is a positive parameter. This model is discussed in Ref.~\cite{Miranda}. The model in Eq.~(\ref{log_model_class_3_1}) reduces to the $\Lambda$CDM model faster than the one in Eq.~(\ref{log_model_class_3_2}).

\vspace{3mm}
\noindent\textbullet{ \emph{Logarithmic format 2:}}
\be f(R)=R-b\left[c-\frac{1}{\log(1+R/R_0)}\right].\ee
Example parameters for this model are $b=R_0=1$ and $c=5$.

\vspace{3mm}
\noindent\textbullet{ \emph{Polynomial format:}}
\be f(R)=R-b\left[c-\left(\frac{R_0}{R}\right)^{n}\right].\label{polynomial_model_class_3_1}\ee
Example parameters for this model are $b=R_0=1$ and $c=n=2$. This model and the Hu-Sawicki model in Eq.~(\ref{Hu_Sawicki_model}) can be considered to be modifications of the $1/R$ model, in which $f(R)=R-\mu^4/R$, where $\mu$ is a parameter with units of mass~\cite{Carroll}. In the $1/R$ model, $f''$ and $V''(\phi)$ are negative, so the scalaron $f'$ is tachyonic. This problem is avoided in the modified versions in Eqs.~(\ref{Hu_Sawicki_model}) and (\ref{polynomial_model_class_3_1}).

\vspace{3mm}
\noindent\textbullet{ \emph{Exponential format 1:}}
\be f(R)=R-b\left[c-\exp\left(-R/R_0\right)\right].\ee
Example parameters for this model are $b=R_0=1$ and $c=2$. This model is explored in Refs.~\cite{Cognola,Linder,Bamba,Elizalde}.

\vspace{3mm}
\noindent\textbullet{ \emph{Exponential format 2:}}
\be f(R)=R-b\left[c-\exp(R_0/R)\right].\ee
Example parameters for this model are $b=5$, $c=2$, and $R_0=1$.

\subsection{Type III $\boldsymbol{\Lambda}$CDM-like $\boldsymbol{f(R)}$ models}
One can combine Eqs.~(\ref{Type_I_LCDM_models}) and (\ref{Type_II_LCDM_models}), and obtain the third type of $\Lambda$CDM-like $f(R)$ models
\be f(R)=R-b\frac{c-A(R/R_0)}{d+A(R/R0)}=R+b\left[1-\frac{c+d}{d+A(R/R0)}\right]. \label{Type_III_LCDM_models}\ee
Some models of this type can be constructed as follows.

\vspace{3mm}
\noindent\textbullet{ \emph{Exponential format:}}
\be
\begin{split}
f(R)& =R-b\tanh(R/R_0)\\
& =R-b\frac{1-\exp(-2R/R_0)}{1+\exp(-2R/R0)}\\
& =R+b\left[1-\frac{2}{1+\exp(-2R/R_0)}\right].
\end{split}
\label{Type_III_tanh}
\ee
Example parameters for this model are $b=R_0=1$. This model is presented in Ref.~\cite{Tsujikawa}. The model described by Eq.~(\ref{Type_I_exp_Oikonomou}) is almost the same as this model.

\vspace{3mm}
\noindent\textbullet{ \emph{Logarithmic format:}}
\be f(R)=R-b\frac{c-\log(1+R_0/R)}{d+\log(1+R_0/R)}.\ee
Example parameters for this model are $b=6$ and $c=d=R_0=1$.

\vspace{3mm}
\noindent\textbullet{ \emph{Polynomial format:}}
\be f(R)=R-b\frac{c-(R_0/R)^{n}}{d+(R_0/R)^{n}}.\ee
Example parameters for this model are $b=6$, $c=d=R_0=1$, and $n=2$.
\section{Cosmological evolution\label{sec:cosmological_evolution}}
In this section, we will explore the cosmological evolution of the $\Lambda$CDM-like $f(R)$ gravity by taking the model in Eq.~(\ref{log_model_class_2_1}) as an example.
\subsection{Formalism}
In this paper, we consider the homogeneous universe in the flat Friedmann-Robertson-Walker metric, $ds^{2}=-dt^{2}+a^{2}(t)d\vec{x}^{2}$. In this case, the universe can be modeled by a four-dimensional dynamical system of $\{\phi,\pi,H,a\}$, where
\be \pi\equiv\dot{\phi}, \label{pi_definition}\ee
$H$ is the Hubble parameter, and the dot $(\cdot)$ denotes the derivative with respect to time. Equation~(\ref{trace_eq1}) provides the dynamical equation for $\pi$,
\be \dot{\pi}=-3H\pi-V'(\phi)+\frac{8\pi G}{3}{\rho_m}. \label{pi_dot}\ee
The equation of motion for $H$ is
\be \dot{H}=\frac{R}{6} -2H^{2}. \label{H_dot} \ee
The definition of the Hubble parameter gives
\be \dot{a}=aH. \label{a_dot}\ee
The system is constrained by
\be H^{2}+\frac{\pi}{\phi}H+\frac{1}{6} \frac{f-\phi R}{\phi}-\frac{8\pi G}{3\phi}(\rho_{m}+\rho_{r})=0,
\label{constraint_eq} \ee
where $\rho_m$ and $\rho_r$ are the densities of matter and radiation, respectively.
Equations~(\ref{pi_definition})-(\ref{constraint_eq}) provide a closed description of the dynamical system of $\{\phi,\pi,H,a\}$.
\begin{figure}
\includegraphics[width=7.5cm,height=5.8cm]{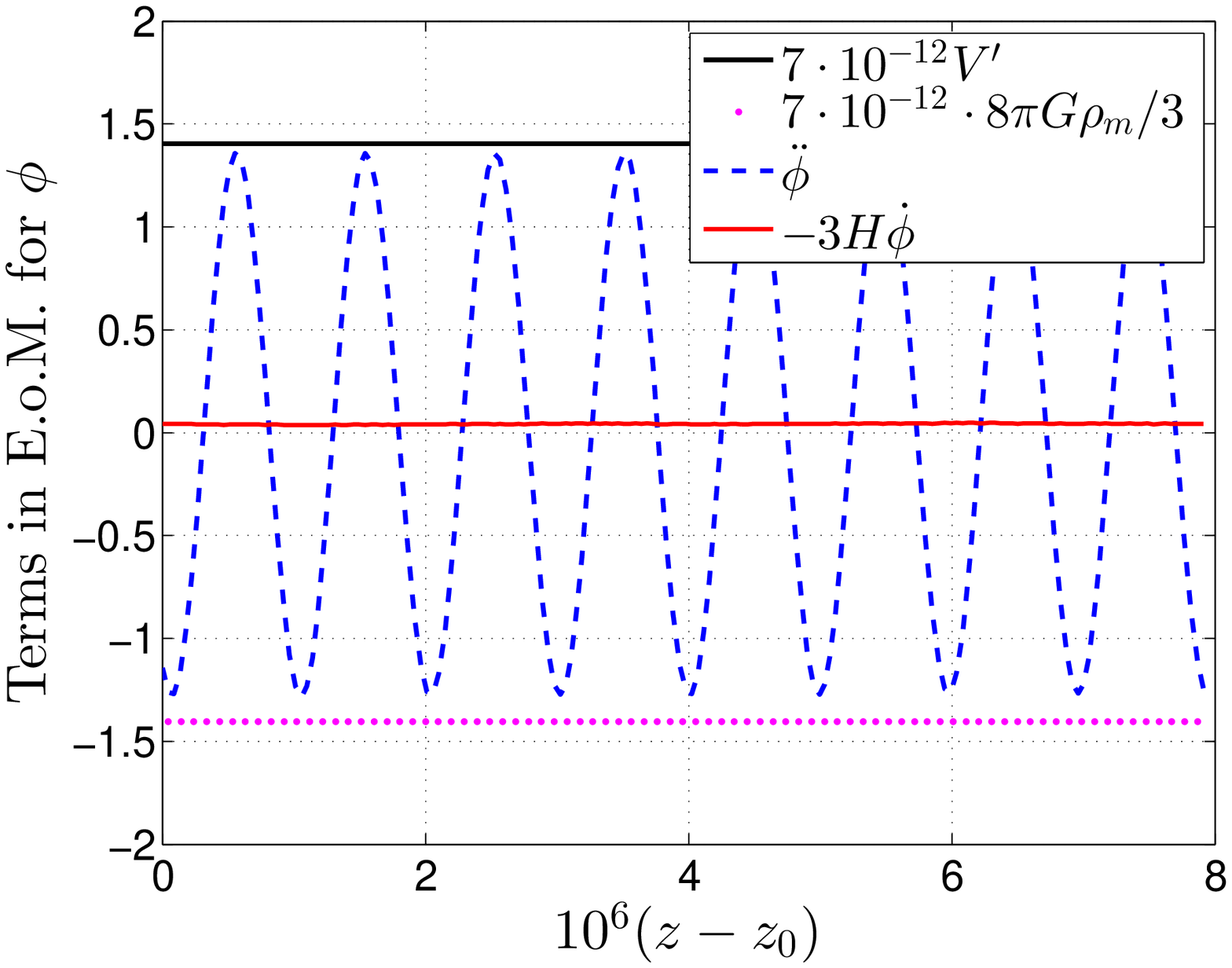}
\caption{The numerical evolution at high redshift for the model of Eq.~(\ref{log_model_class_2_1}) with $b=1.5$ and $c=R_0=1$. $z_{0}=20256.709810835$.
At high redshift, the field $\phi$ evolves very slowly and oscillates between $V'(\phi)$ and $8\pi G\rho_m/3$. The figure shows the relation between the four terms in Eq.~(\ref{pi_dot}): $|3H\dot{\phi}|<|\ddot{\phi}|\ll V'(\phi)\approx 8\pi G\rho_m/3$.}
\label{fig:oscillations_phi_dubdot}
%
\includegraphics[width=7.5cm,height=5.8cm]{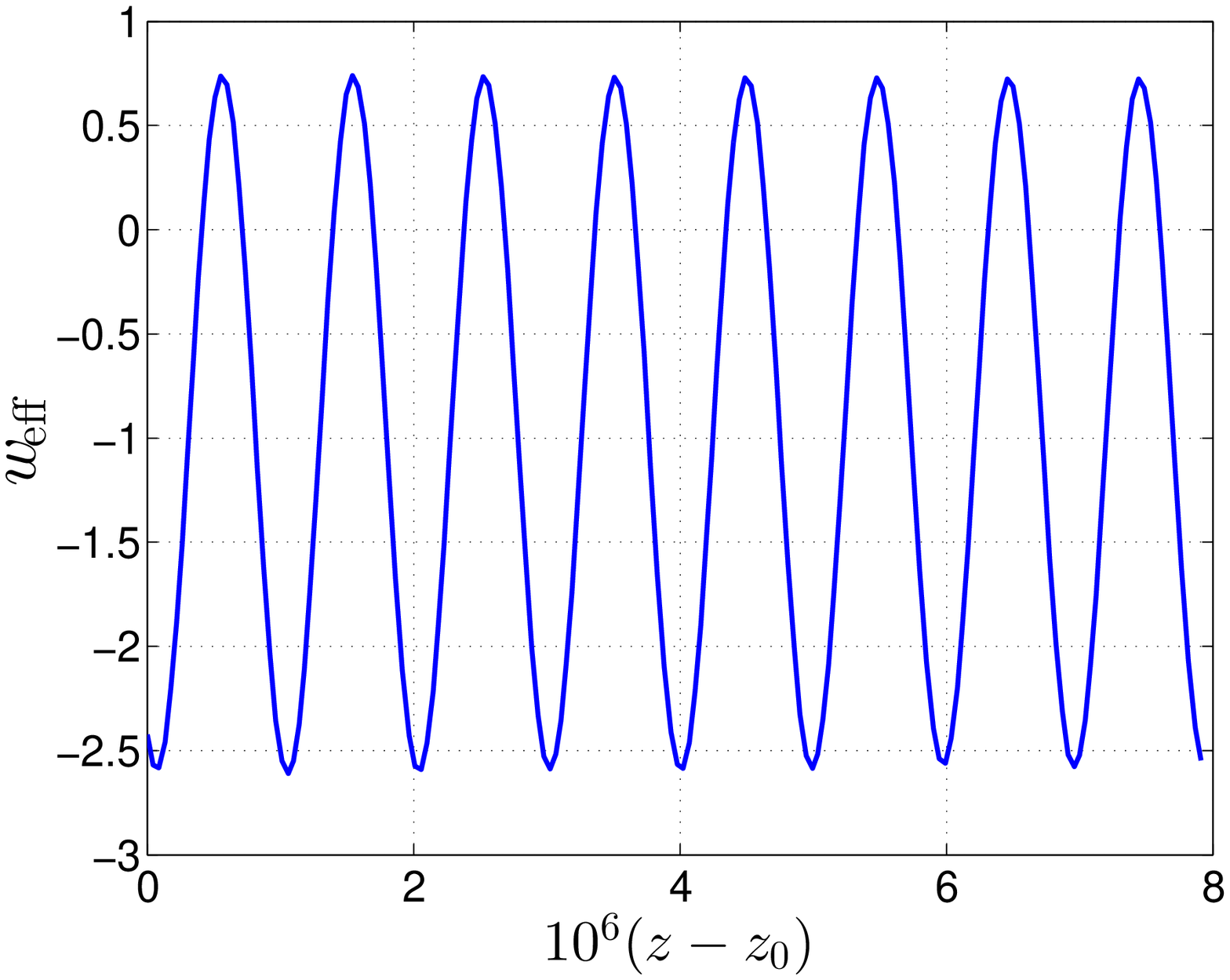}
\caption{The oscillations of $w_{\mbox{eff}}$ in the numerical evolution at high redshift for the model of Eq.~(\ref{log_model_class_2_1}) with $b=1.5$ and $c=R_0=1$. $z_{0}=20256.709810835$. As a consequence of the oscillations of the field $\phi$, the equation of state $w_{\text{eff}}$ also oscillates near $-1$.}
\label{fig:oscillations_w_eff}
\end{figure}
\subsection{The evolution from the exact method\label{sec:evolution_exact_method}}
\begin{figure}
\includegraphics[width=7.5cm,height=5.8cm]{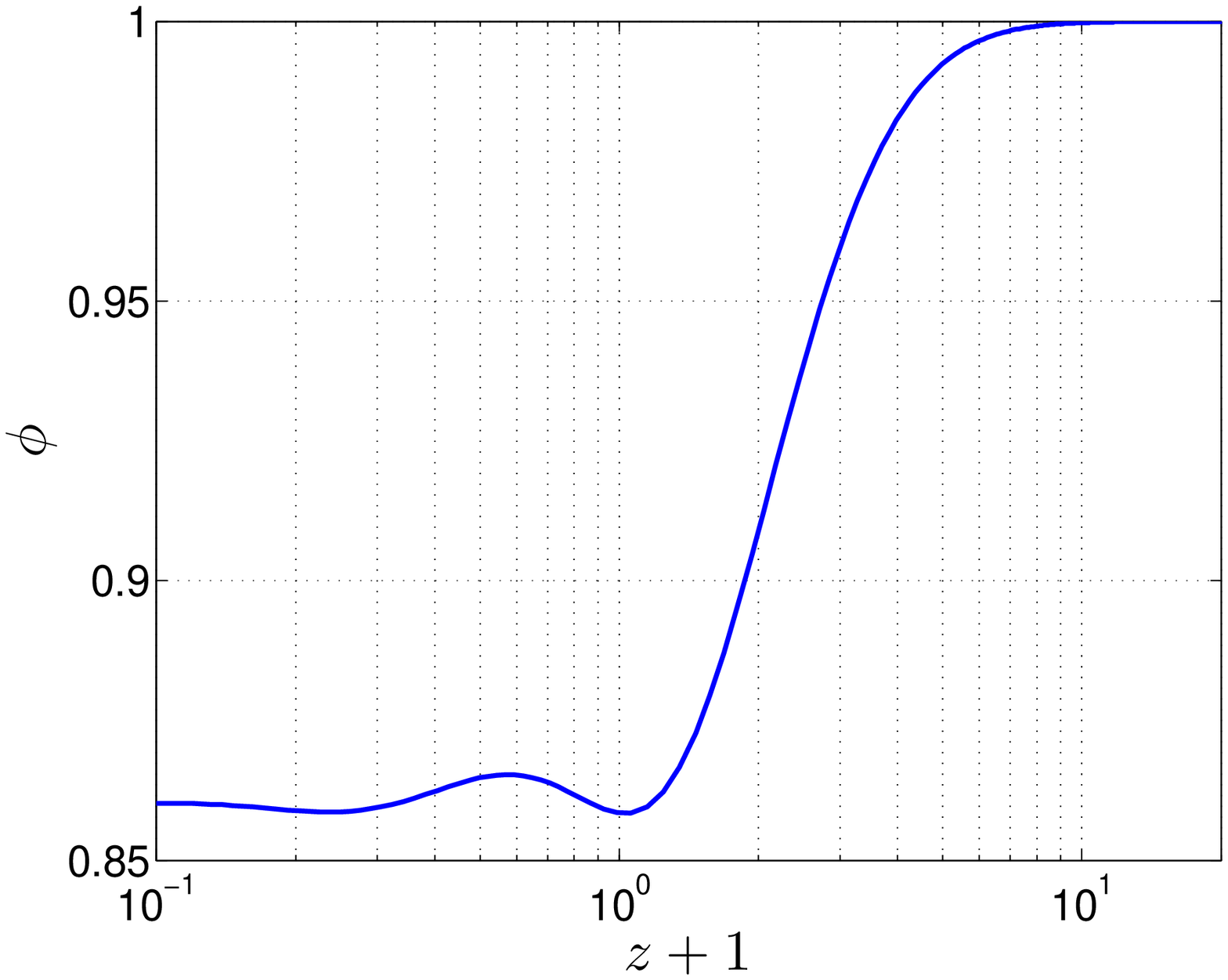}
\caption{The evolution for $\phi$ at low redshift for the model of Eq.~(\ref{log_model_class_2_1}) with $b=1.5$ and $c=R_0=1$. The field $\phi$ has a slow roll when $8\pi G\rho_m>\Lambda$. After that, the field drops significantly, oscillates, and eventually stops at the minimum of the potential $V(\phi)$ due to friction force $-3H\dot{\phi}$.}
\label{fig:phi_low_z}
%
\includegraphics[width=7.5cm,height=5.8cm]{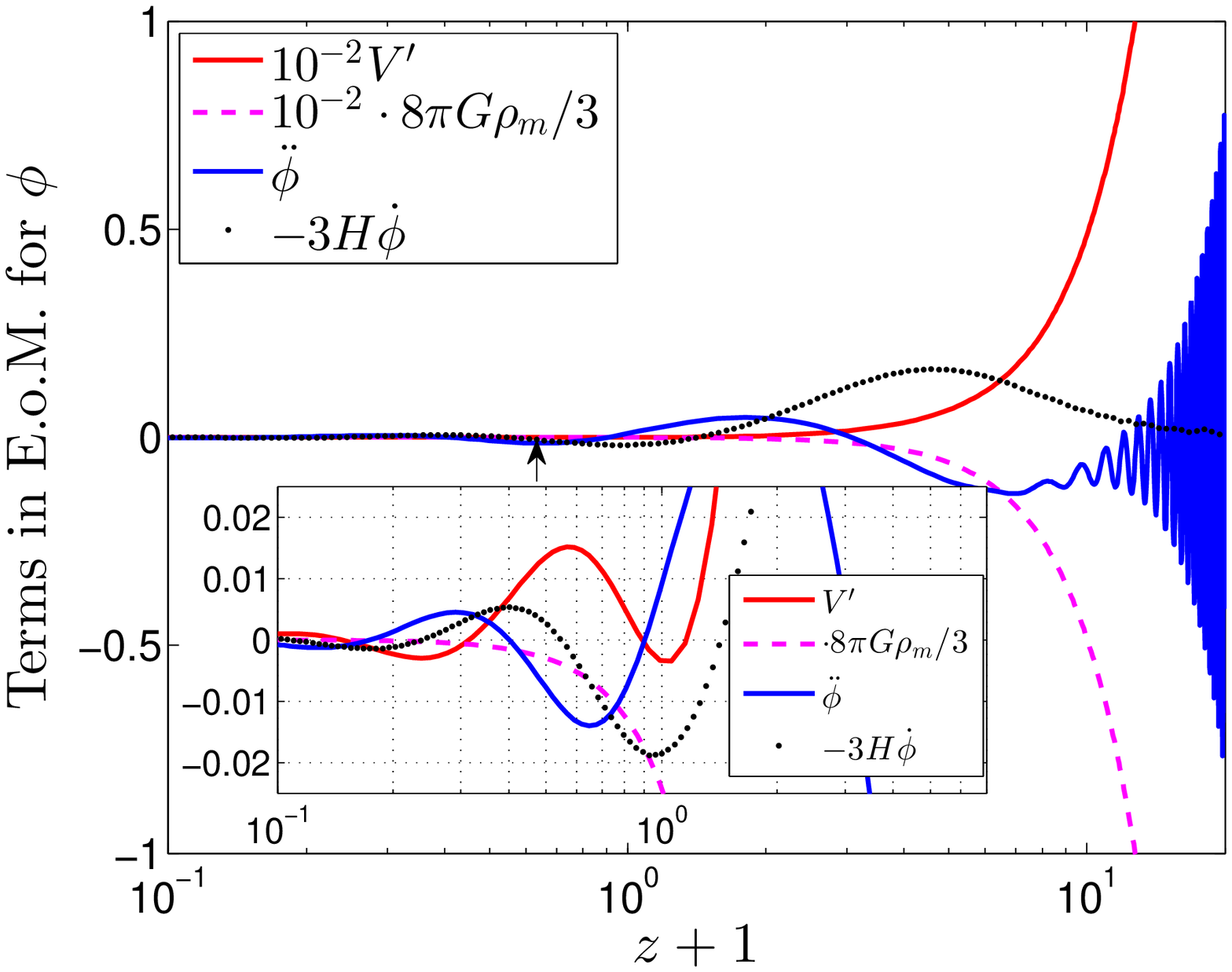}
\caption{The terms in the equation of motion (E.o.M.) (\ref{pi_dot}) for $\phi$ at low redshift for the model of Eq.~(\ref{log_model_class_2_1}) with $b=1.5$ and $c=R_0=1$. In the late universe, compared to other terms in Eq.~(\ref{pi_dot}), the matter force term $8\pi G\rho_m/3$ is negligible, and Eq.~(\ref{pi_dot}) is reduced to: $\ddot{\phi}\approx-3H\dot{\phi}-V'(\phi)$.}
\label{fig:EoM_low_z}
\end{figure}
\begin{figure}
\includegraphics[width=7.5cm,height=5.8cm]{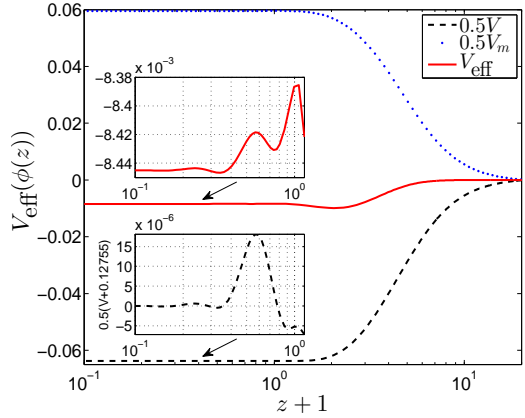}
\caption{$V_{\text{eff}}(\phi)$ vs. $z+1$ at low redshift for the model of Eq.~(\ref{log_model_class_2_1}) with $b=1.5$ and $c=R_0=1$.
The potentials $V$, $V_m$, and $V_{\text{eff}}$ are defined by Eqs.~(\ref{v_prime}), (\ref{V_eff_definition}), and (\ref{V_m_definition}), respectively. $V_{\text{eff}}=V+V_m$. $V_{\text{eff}}$ is very flat at high redshift and has a drop at low redshift, such that the field $\phi$ has a slow roll at high redshift and then a drop at low redshift.}
\label{fig:V_eff_low_z}
\end{figure}
\begin{figure}
\includegraphics[width=7.5cm,height=5.8cm]{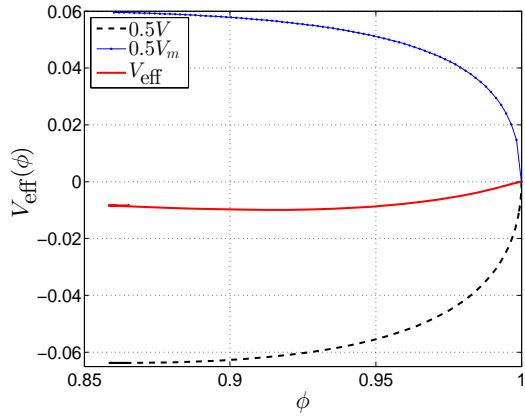}
\caption{$V_{\text{eff}}(\phi)$ vs. $\phi$ at low redshift for the model of Eq.~(\ref{log_model_class_2_1}) with $b=1.5$ and $c=R_0=1$.
$V_{\text{eff}}=V+V_m$.}
\label{fig:V_eff_phi_low_z}
\end{figure}
\begin{figure}
\includegraphics[width=7.5cm,height=5.8cm]{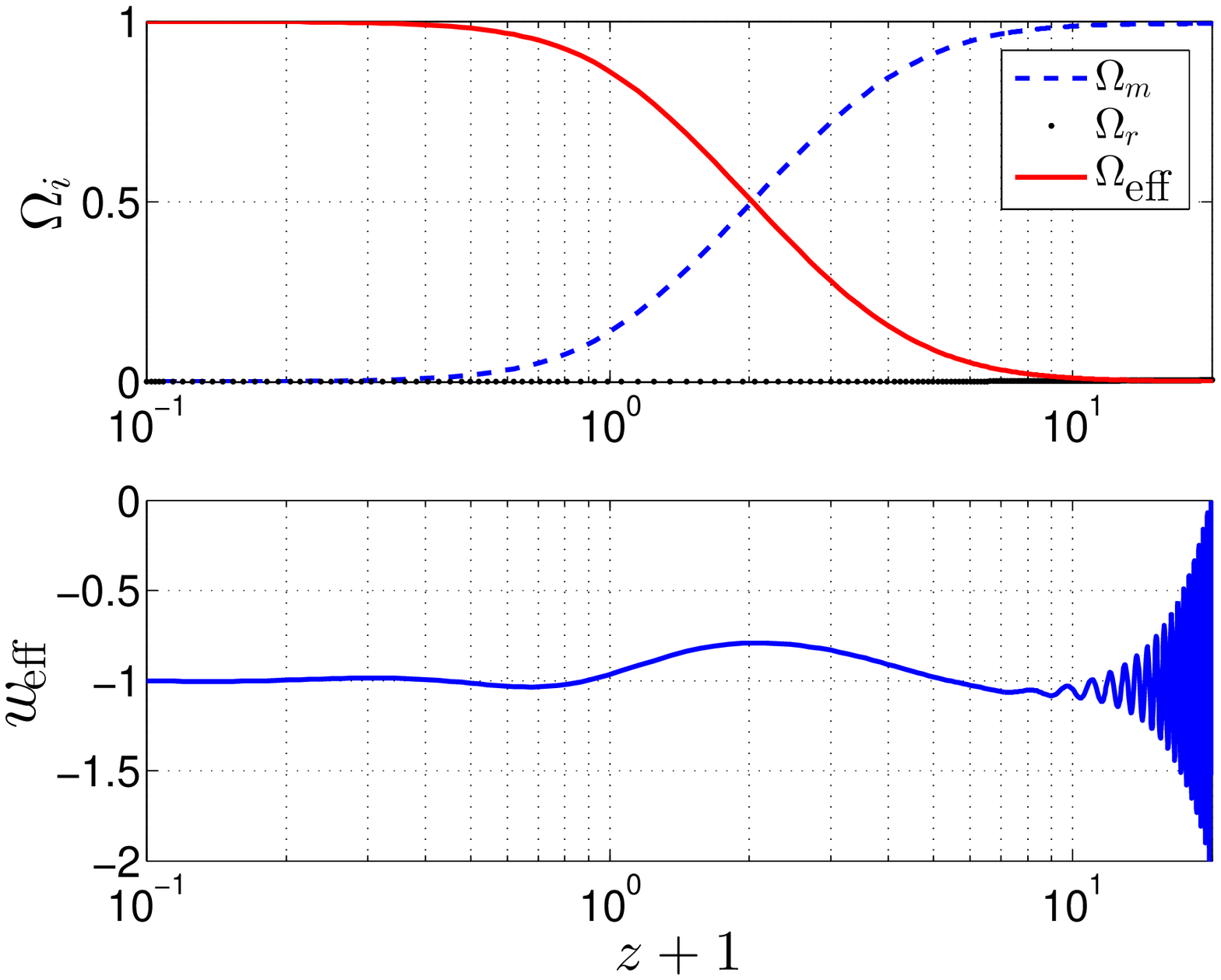}
\caption{The evolutions of ${\Omega_i}'s$ and $w_{\text{eff}}$ at low redshift for the model of Eq.~(\ref{log_model_class_2_1}) with $b=1.5$ and $c=R_0=1$.} \label{fig:evolution_exact_low_z}
\end{figure}
A straightforward way to simulate cosmological evolution is to integrate the equations of motion (\ref{pi_dot})-(\ref{a_dot}). However, from the point of view of numerics, at high redshift, the field $\phi$ evolves very slowly and oscillates between $V'(\phi)$ and $8\pi G\rho_m/3$. The behaviors of the terms in Eq.~(\ref{pi_dot}) are shown in Fig.~\ref{fig:oscillations_phi_dubdot}, revealing the relation between the four terms in Eq.~(\ref{pi_dot}):
$|3H\dot{\phi}|<|\ddot{\phi}|\ll V'(\phi)\approx 8\pi G\rho_m/3$. The field $\phi$ oscillates near the minimum of the effective
potential $V_{\text{eff}}(\phi)$, which is defined by
\be V_{\text{eff}}'(\phi)=V'(\phi)+V_{m}'(\phi), \label{V_eff_definition}\ee
with
\be V_{m}'(\phi)=-8\pi G\rho_m/3. \label{V_m_definition}\ee
These oscillations produce particles, and could be a possible source of energetic cosmic rays~\cite{Arbuzova}. The oscillations of $\phi$ also make it inconvenient to numerically integrate the evolution in the early universe. Therefore, in the next sub-section, we consider employing an approximation method instead. As a consequence of the oscillations of the field $\phi$, the equation of state $w_{\text{eff}}$ also oscillates near $-1$, as shown in Fig.~\ref{fig:oscillations_w_eff}.

The behaviors of the field $\phi$ and the terms in the equation of motion for $\phi$ (\ref{pi_dot}) at low redshift are shown in Figs.~\ref{fig:phi_low_z} and \ref{fig:EoM_low_z}, respectively. Figure~\ref{fig:phi_low_z} reveals that the field $\phi$ has a slow roll when the matter density is greater than the cosmological constant. The field $\phi$ drops significantly as the matter density comes to the cosmological constant scale, oscillates at the minimum of the potential $V(\phi)$, and then eventually stops due to friction force $-3H\dot{\phi}$. As demonstrated in Fig.~\ref{fig:EoM_low_z}, in the late universe, compared to other terms in Eq.~(\ref{pi_dot}), the matter force term $8\pi G\rho_m/3$ is negligible, and Eq.~(\ref{pi_dot}) is reduced to
\be \ddot{\phi}\approx-3H\dot{\phi}-V'(\phi).\ee
These results can also be interpreted via the effective potential $V_{\text{eff}}(\phi)$ as plotted in Figs.~\ref{fig:V_eff_low_z} and \ref{fig:V_eff_phi_low_z}. The effective potential $V_{\text{eff}}$ is very flat at high redshift and has a drop at low redshift, such that the field $\phi$ has a slow roll at high redshift and then a drop at low redshift. The evolutions of ${\Omega_i}'s$ and $w_{\text{eff}}$ at low redshift are shown in Fig.~\ref{fig:evolution_exact_low_z}. The ${\Omega_i}'s$ are defined as $\Omega_{i}=8\pi G\rho_{i}/(3H^2)$, where $i$ refers to the indexing of radiation, matter or effective dark energy. The $\Lambda$CDM-like models can mimic a cosmological evolution, fitting the observations without difficulty. Comparison of Figs.~\ref{fig:oscillations_w_eff} and \ref{fig:evolution_exact_low_z} demonstrates that the phantom behavior of $w_{\text{eff}}$ ($w_{\text{eff}}$ crosses $-1$) at low redshift is nothing but an extension of that behavior at high redshift.

\subsection{The evolution from the approximation method}
\begin{figure}
\includegraphics[width=7.5cm,height=5.8cm]{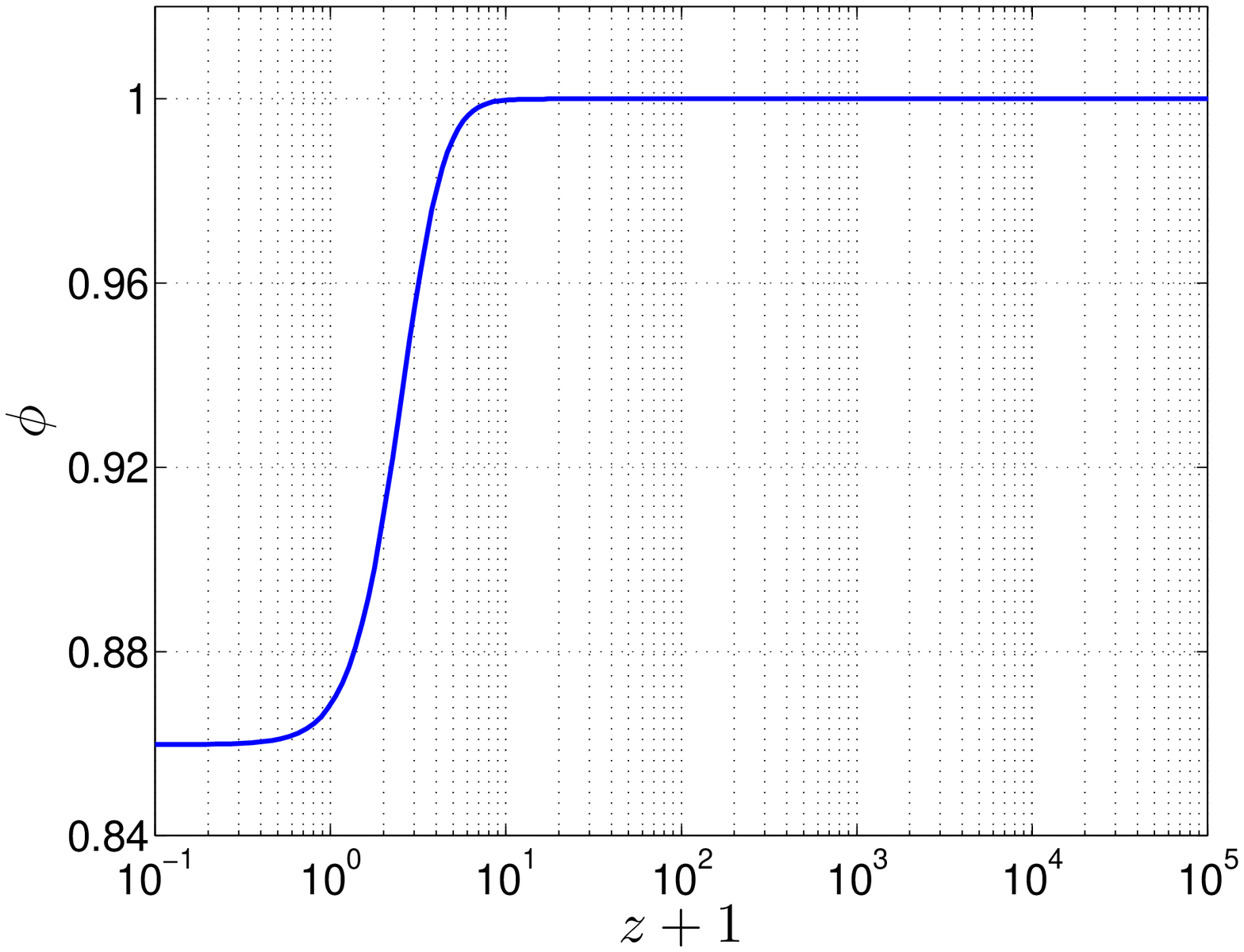}
\caption{The evolution for $\phi$ in the approximation method for the model of Eq.~(\ref{log_model_class_2_1}) with $b=1.5$ and $c=R_0=1$. The field $\phi$ has a slow roll at high redshift and a drop at low redshift.} \label{fig:phi_approximate}
%
\includegraphics[width=7.5cm,height=5.8cm]{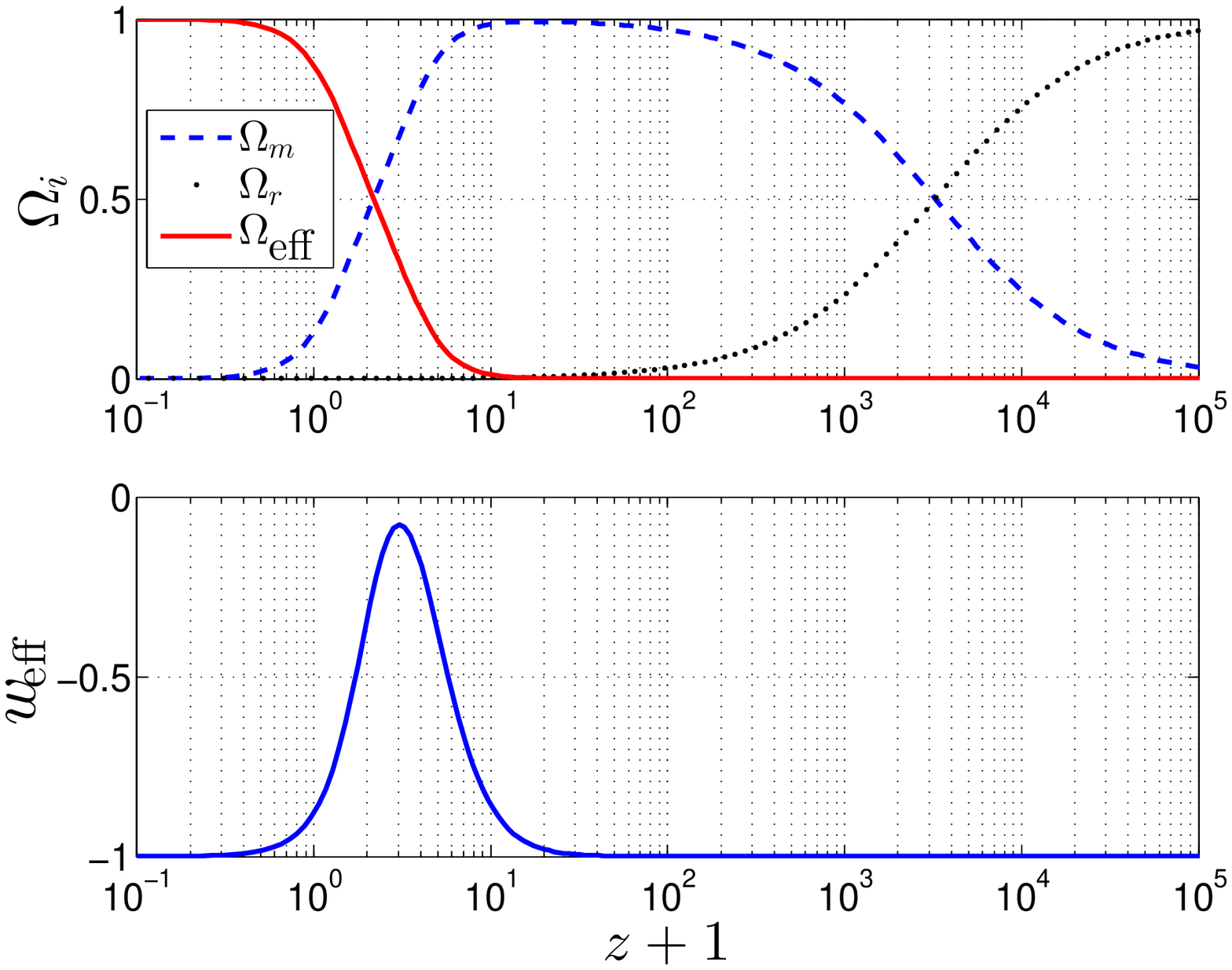}
\caption{The evolutions of ${\Omega_i}'s$ and $w_{\text{eff}}$ in the approximation method for the model of Eq.~(\ref{log_model_class_2_1}) with $b=1.5$ and $c=R_0=1$.}
\label{fig:Omega_approximate}
\end{figure}
\begin{figure}
\includegraphics[width=7.5cm,height=5.8cm]{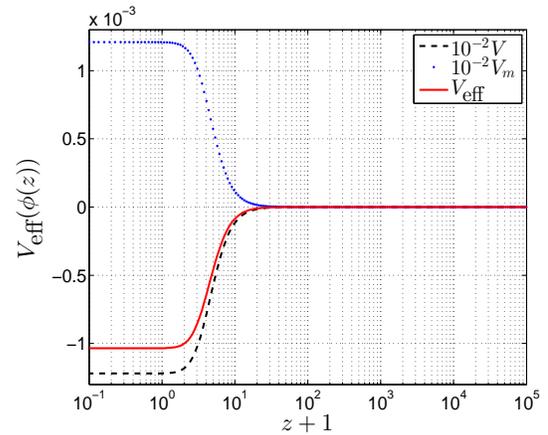}
\caption{The effective potential $V_{\text{eff}}(\phi)$ in the approximation method for the model of Eq.~(\ref{log_model_class_2_1}) with $b=1.5$ and $c=R_0=1$. The potentials $V$, $V_m$, and $V_{\text{eff}}$ are defined by Eqs.~(\ref{v_prime}), (\ref{V_eff_definition}), and (\ref{V_m_definition}), respectively. $V_{\text{eff}}=V+V_m$. $V_{\text{eff}}$ is very flat at high redshift and has a drop at low redshift, such that the field $\phi$ has a slow roll at high redshift and a drop at low redshift.}
\label{fig:potential_approximate}
\end{figure}
Due to the oscillations of $\phi$, it is not convenient to numerically integrate the cosmological evolution in the early universe. However, since the field $\phi$ oscillates near the minimum of the effective potential $V_{\text{eff}}(\phi)$ in the early universe, one can take the evolution of the minimum of $V_{\text{eff}}(\phi)$ approximately to be that of the field $\phi$. The approximate solution describes the evolution of $\phi$ accurately in the early universe. In the late universe, the deviation of the field $\phi$ from the minimum of $V_{\text{eff}}(\phi)$ becomes relatively large, and the exact method should be used as has been done in Sec.~\ref{sec:evolution_exact_method}.

Equations~(\ref{pi_definition}) and (\ref{pi_dot}) can be combined as
\be \ddot{\phi}=-3H\dot{\phi}-V'(\phi)+\frac{8\pi G}{3}{\rho_m}. \ee
For the $\Lambda$CDM-like models, the field $\phi$ evolves very slowly in the early universe. From this we get
\be |3H\dot{\phi}| < |\ddot{\phi}| \ll V'(\phi)\approx \frac{8\pi G}{3}{\rho_m},\label{v_rho_dynamical} \ee
as shown in Fig.~\ref{fig:oscillations_phi_dubdot}. Note that $\rho_m=\rho_{m0}/a^3$ and $\dot{a}=aH$, where $\rho_{m0}$ is the matter density of the current universe. These, together with $V'(\phi)\approx 8\pi G\rho_m/3$, lead to
\be \dot{\phi}\approx-3H\frac{V'}{V''}. \label{phi_dot_approximate}\ee
Therefore, in our approximation method, we take Eq.~(\ref{phi_dot_approximate}) as an approximate equation of motion for $\phi$, we replace Eqs.~(\ref{pi_definition}) and (\ref{pi_dot}) by Eq.~(\ref{phi_dot_approximate}), but keep Eqs.~(\ref{H_dot})-(\ref{constraint_eq}).
The results of the approximation method are shown in Figs.~\ref{fig:phi_approximate}-\ref{fig:potential_approximate}. The field $\phi$ has a slow roll evolution in the early universe due to the approximate balance between $V'$ and $8\pi G\rho_m/3$, and the the evolutions of ${\Omega_i}'s$ are consistent with cosmological observations.

The exact and the approximate methods are supplementary. The numerical integration in the exact method at high redshift is very slow due to the oscillations of the field $f'$, but can mimic the cosmological evolution in the late universe very easily. The approximate method does not yield an accurate cosmological evolution in the late universe, but can mimic a smooth evolution in the early universe. Therefore, these two methods can be used together to explore the cosmic dynamics of $f(R)$ gravity.

\subsection{The equation of state $\boldsymbol{w_{\mbox{eff}}}$}
\begin{figure}
\includegraphics[width=7.5cm,height=5.8cm]{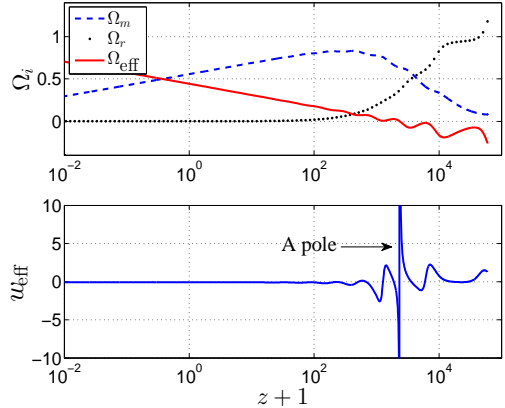}
\caption{The equation of state $w_{\text{eff}}$ for the $R\ln R$ model with $\alpha=0.02$ and $R_0=1$. Since $f'$ is greater than 1 in the early universe, the effective dark-energy density $\rho_{\text{eff}}$ described by Eq.~(\ref{rho_eff}) can be zero at some moment, which will cause a pole in the equation of state $w_{\text{eff}}$. } \label{fig:pole_w_eff_evolution_RlnR}
\end{figure}

The Big Bang nucleosynthesis and the observations of the Cosmic Microwave Background imply that general relativity should be recovered in the early universe, which means that $f(R)\rightarrow R$ and $f'\rightarrow 1$ as $R\gg \Lambda$. On the other hand, $f''$ should be positive to ensure that the scalaron $f'$ is non-tachyonic. Consequently, $f'$ should be less than 1~\cite{Pogosian}. In fact, an $f'$ greater than 1 in the early universe can cause a pole in the equation of state $w_{\text{eff}}$, as shown in Fig.~\ref{fig:pole_w_eff_evolution_RlnR} and also pointed out in Ref.~\cite{Frolov}. This situation can be explained with Eqs.~(\ref{w_eff}) and (\ref{rho_eff}). The first term, $(f'R-f)/2$, in Eq.~(\ref{rho_eff}) can be positive. (This term is equal to $\alpha R/2$ and is positive for the $R\ln R$ model, in which $f(R)=R[1+\alpha \ln (R/R_0)]$ and $\alpha$ is positive.) The second term, $-3H\dot{f'}$, is positive because $f'$ will roll down to the minimum of the potential $V(\phi)$. However, the third term, $3H^{2}(1-f')$, is negative when $f'$ is greater than 1 in the early universe. Then, in the later evolution, $f'$ will decrease, cross 1, and move to the minimum of the potential $V(\phi)$. At some moment, the energy density of the effective dark energy $\rho_{\text{eff}}$ will be zero. This will generate a pole in the equation of state $w_{\text{eff}}$ defined by Eq.~(\ref{w_eff}). The $\Lambda$CDM-like models do not have such a problem, since the function $f(R)\approx R-2\Lambda$ and $f'\approx 1$ in the early universe.

A feasible $f(R)$ model should also pass the solar system tests. We studied how the Hu-Sawicki model deals with the solar system tests in Ref.~\cite{Guo2}. It turned out that the $\Lambda$CDM-like $f(R)$ models have the advantage of passing the solar system tests.
\\
\section{Conclusions\label{sec:conclusions}}
In this paper, we studied how to construct $f(R)$ gravity models which satisfy the stability and viability conditions. Cosmological observations and local gravity tests place stringent requirements on the format of the function $f(R)$. A feasible $f(R)$ model needs to be very close to the $\Lambda$CDM model. For the $\Lambda$CDM-like models, it is not hard to obtain the recovery of general relativity in the early universe and a cosmic speed-up in the late universe. The condition $f'>0$ can also be easily satisfied since the modification term is usually small compared to the main term $R$ in the function $f(R)$. The thing is how to make the potential $V(\phi)$ have a minimum. One way is to tune the parameters, such that $V''(\phi)>0$, $V'(\phi)>0$ when $R\gg \Lambda$, and $V'(\phi)<0$ when $R<\Lambda$. Once these have been achieved, the potential $V(\phi)$ will have a minimum. With this method, three types of $\Lambda$CDM-like $f(R)$ models expressed by Eqs.~(\ref{Type_I_LCDM_models}), (\ref{Type_II_LCDM_models}), and (\ref{Type_III_LCDM_models}) are constructed. In addition to these three types, a viable $\Lambda$CDM-like $f(R)$ may also take other forms. For example, the Starobinsky model
takes the form $f(R)=R+\lambda R_0[(1+R^2/R_0^2)^{-n}-1]$~\cite{Starobinsky}, and a new exponential model takes
the form $f(R)=(R-\lambda c)\exp[\lambda(c/R)^{n}]$ ~\cite{Xu}.

We also studied the cosmological evolution of the $\Lambda$CDM-like $f(R)$ models. The field $\phi$ evolves slowly in the early universe due to the quasi-static balance between $V'(\phi)$ and $8\pi G\rho_m/3$, and is released from the coupling between $V'(\phi)$ and $8\pi G\rho_m/3$ in the late universe to generate a dark-energy domination epoch. The numerical simulation is slow in the early universe because of the oscillations of the field $\phi$ near the minimum of the effective potential $V_{\text{eff}}(\phi)$. To avoid this problem, we take the minimum of $V_{\text{eff}}(\phi)$ as an approximate solution for $\phi$ and obtain the cosmological evolution from the early universe to the late one. This approximation method describes the cosmological evolution well except in the late universe when the curvature scalar is below the cosmological constant scale. We use the exact method to study the late-universe evolution. Then, a combination of the exact and the approximate methods provides a complete picture of the cosmological evolution of $f(R)$ gravity.

\section*{Acknowledgments}\small
This  work  was  supported  by  the  Discovery  Grants  program  of  the  Natural  Sciences  and
Engineering Research Council of Canada. The author would  like  to thank Andrei V. Frolov and Levon Pogosian for useful discussions.


\end{document}